\documentclass{trbunofficial}
\usepackage{graphicx}
\usepackage{amsmath}
\usepackage{mathtools}
\usepackage{subcaption}
\usepackage{bm}
\usepackage{amsfonts}

\usepackage[hidelinks]{hyperref}

\usepackage{booktabs, tabularx}

\graphicspath{ {./images/} }

\AuthorHeaders{Zhang, LaRock, Bagabaldo, and Hackl}
\title{Rethinking the Sioux Falls Network: Insights from Path-Driven Higher-Order Network Analysis}

\author{%
  \textbf{Chen Zhang}\\
  Department of Civil and Environmental Engineering\\
  Princeton University, Princeton, NJ 08544, United States\\
  Email: cz2687@princeton.edu\\
  \hfill\break
  \textbf{Timothy LaRock, Ph.D.}\\
  Department of Civil and Environmental Engineering\\
  Princeton University, Princeton, NJ 08544, United States\\
  Email: larock@princeton.edu\\
  \hfill\break%
  \textbf{Alben Rome Bagabaldo, Ph.D.}\\
  Department of Civil and Environmental Engineering\\
  Princeton University, Princeton, NJ 08544, United States\\
  Email: alben@princeton.edu\\
  \hfill\break%
  \textbf{Jürgen Hackl, Ph.D., Corresponding Author}\\
  Department of Civil and Environmental Engineering\\
  Princeton University, Princeton, NJ 08544, United States\\
  Email: hackl@princeton.edu
}

\WordsPerTable{250}

\TotalWords{7157}

\begin{document}
\maketitle

\section{Abstract}
Benchmark scenarios are widely used in transportation research to evaluate routing algorithms, simulate infrastructure interventions, and test new technologies under controlled conditions.
However, the structural and behavioral fidelity of these benchmarks remains largely unquantified, raising concerns about the external validity of simulation results.
In this study, we introduce a mathematical framework based on higher-order network models to evaluate the representativeness of benchmark networks, focusing on the widely used Sioux Falls scenario.
Higher-order network models encode empirical and simulated trajectory data into memory-aware network representations, which we use to quantify sequential dependencies in mobility behavior and assess how well benchmark networks capture real-world structural and functional patterns.
Applying this framework to the Sioux Falls network, as well as real-world trajectory data, we quantify structural complexity, optimal memory length, link prediction accuracy, and centrality alignment.
Our results show and statistically quantify that the classical Sioux Falls network exhibits limited path diversity, rapid structural fragmentation at higher orders, and weak alignment with empirical routing behavior.
These results illustrate the potential of higher-order network models to bridge the gap between simulation-based and real-world mobility analysis, providing a robust foundation for more accurate and generalizable insights in transportation research.

\hfill\break%
\noindent\textit{Keywords}: Higher-Order Networks, Benchmark, Sioux Falls, Stochastic Network Analysis, Model Validation, Agent-Based Simulation

\newpage

\section{Introduction}
%
%
Transportation networks are a critical foundation of modern society, enabling the movement of people, goods, and services across urban and regional scales.
To meet these demands, they function as complex, adaptive systems shaped by physical infrastructure, individual choices, and emergent collective dynamics.
Transportation networks are uniquely complex systems due to their large spatial scales, integration with the physical environment, and the bidirectional influence between infrastructure and user behavior.
Because large-scale field testing is often impractical or infeasible, the design and evaluation of transportation systems typically rely on simulation and virtual prototyping \cite{Ungureanu2018Civil}.
To understand and manage traffic flow within these complex systems, researchers model network dynamics using a wide range of techniques drawn from nonequilibrium statistical physics and nonlinear dynamics \cite{Mahnke2005Probabilistic}.
These approaches span from large-scale \textit{macroscopic} models, where the traffic movement resembles an incompressible fluid \cite{Wang2022Macroscopic}, towards \textit{microscopic} models, where the focus shifts to the interactions between individuals \cite{Nguyen2021overview}. 

%
In the absence of large-scale experimental validation, benchmark scenarios and virtual testbeds have become indispensable for studying, evaluating, and improving transportation networks.
These controlled, reproducible environments allow researchers to systematically explore complex system behavior and test the impact of new technologies or interventions under varying conditions.
Benchmark networks are widely used in applications such as traffic assignment \cite{Cantarella2019Solving}, route guidance optimization \cite{Movaghar2025Joint}, evacuation planning \cite{Xu2024Exploring}, infrastructure investment analysis \cite{Nakazato2024365day}, and the deployment of connected and autonomous vehicles \cite{Mirzahossein2024Enhancing}.
They also serve as critical platforms for developing and validating agent-based simulations \cite{Hackl2019Epidemic}, travel demand models \cite{Chakirov2014Enriched}, and network design strategies \cite{Baskan2014Harmony}, especially when empirical data is limited or ethically challenging to obtain in the real world \cite{Hackl2019Estimation}.

%
A prominent example of such a benchmark is the Sioux Falls network, introduced by Morlok et al. \cite{Morlok1973Development} and LeBlanc et al. \cite{LeBlanc1975Algorithm, LeBlanc1975efficient}, which has been widely adopted as a testbed in transportation research.
With 24 nodes and 76 links, its compact size offers a manageable environment for evaluating a wide range of algorithms and models (see Section~\ref{sec:background}) \cite{Stabler2025Transportation}.
However, analyses based exclusively on such simplified benchmarks risk producing results that are not consistent with or generalizable to more complex urban networks or empirically observed mobility patterns.
This concern is not restricted to the Sioux Falls network alone, but reflects a broader methodological limitation associated with the use of stylized benchmark scenarios in transportation science.
In the absence of systematic validation, it remains unclear to what extent these networks precisely capture the structural, functional, and behavioral complexities inherent in real-world systems.
Consequently, the external validity of findings derived from benchmark-based simulations is difficult to ascertain, limiting their applicability to practical planning, policy, and design contexts.
Addressing this gap necessitates rigorous frameworks for evaluating the representativeness of benchmark networks and assessing how their structure affects the interpretation and generalization of simulation results.

%
In this study, we introduce a network science approach rooted in statistical physics to evaluate the limitations of benchmark-based simulation studies in transportation research.
Specifically, we leverage higher-order network models \cite{Benson2016Higherorder, Lambiotte2019networks} to construct a mathematically rigorous framework that captures the memory-dependent, path-based nature of mobility behavior \cite{Zhang2025Connectivity}.
This allows us to construct network representations that preserve the statistical properties of empirical path data and bridge the gap between idealized simulations and observed mobility patterns.
Our approach enables a statistically grounded comparison across synthetic benchmarks, simulations of varying scale and complexity, and empirical trajectories.
We demonstrate that higher-order networks not only encode more realistic movement patterns but also serve as a unifying representation to quantify structural and functional deviations between model-based and data-driven network flows.
This framework provides a systematic method for validating simulation outputs against empirical ground truth, quantifying discrepancies in routing behavior, and assessing the representativeness of benchmark networks \cite{Scholtes2017When}.
By applying this methodology to the widely used Sioux Falls network and trajectory data generated through agent-based simulations, we illustrate how higher-order network models can reveal hidden biases and structural artifacts that remain undetected under conventional first-order analyses.
Our contribution thus offers a novel formal framework for evaluating transportation models, enabling a systematic assessment of their external validity and the structural fidelity of benchmark networks relative to empirically observed mobility behavior.
Specifically, this work advances the state-of-the-art in the field transportation research as follows.\\
\begin{itemize}
\item We introduce a \textit{statistically grounded framework} for evaluating the representativeness of benchmark transportation networks by embedding trajectory data into higher-order network models, thereby integrating principles from stochastic network analysis to quantify structural variability, test sequential dependencies, and evaluate model fidelity using formal hypothesis testing and ensemble-based comparisons.

\item We conduct a multi-scale structural and functional \textit{comparison of empirical and simulated mobility systems} using higher-order network analytics, including density evolution, diameter scaling, connected component fragmentation, and degree distribution divergence quantified via Kullback-Leibler divergence and cosine similarity in a multi-order feature space.

\item We demonstrate that benchmark networks, particularly the classical Sioux Falls scenario, exhibit \textit{statistically significant deviations} in path-dependent centrality and predictive accuracy when compared to real-world mobility traces, as revealed through Kendall's Tau correlation of higher-order PageRank rankings and link prediction performance across multiple memory orders.\\
\end{itemize}

%
The remainder of this paper is organized as follows. In Section~\ref{sec:background}, we provide a comprehensive review of benchmark networks in transportation science, with a particular focus on the Sioux Falls scenario, and introduce the principles of stochastic network analysis and higher-order network modeling.
Section~\ref{sec:method} introduces our methodological framework, including the construction of higher-order networks from trajectory data, optimal memory order selection using statistical model comparison, and analytical tasks such as centrality alignment and link prediction.
In Section~\ref{sec:results}, we present a comparative analysis of higher-order structural properties, centrality measures, and predictive performance across real-world and simulated datasets, quantifying the fidelity of benchmark networks in representing empirical mobility behavior.
Section~\ref{sec:discussion} reflects on the implications of our findings for benchmark design, model validation, and generalizability, and discusses the broader relevance of higher-order models for transportation science.
Finally, Section~\ref{sec:conclusion} summarizes the contributions of this study and outlines future research directions, including the integration of multimodal data and behavioral heterogeneity into higher-order mobility models.

\newpage

\section{Background}{\label{sec:background}}

\subsection{Benchmarking in Transportation Science}
%
The advancement of transportation science, like other computational disciplines, depends on rigorous, reproducible, and comparable evaluations of new models and algorithms.
As systems grow in complexity, standardized evaluation frameworks have become essential.
To this end, the field increasingly relies on benchmark scenarios, \emph{i.e.} well-defined, publicly available datasets that specify network topology, link attributes, and demand patterns.
These benchmarks enable direct comparison of methods under identical conditions, facilitating assessment of accuracy, scalability, and efficiency.
The Sioux Falls network is the archetypal example of such a tool.

\begin{figure}[htbp]
  \vspace{-2mm}
  \centering
  \includegraphics[height=4.3cm]{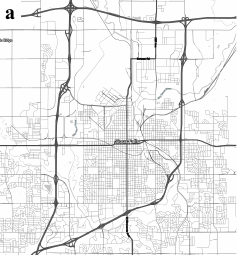}\hfill
  \includegraphics[height=4.3cm]{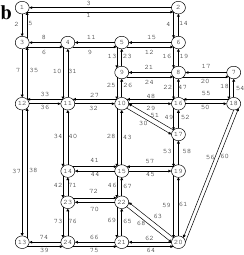}\hfill
  \includegraphics[height=4.3cm]{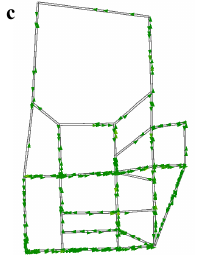}\hfill
  \includegraphics[height=4.3cm]{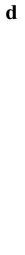}%
  \includegraphics[height=4.3cm]{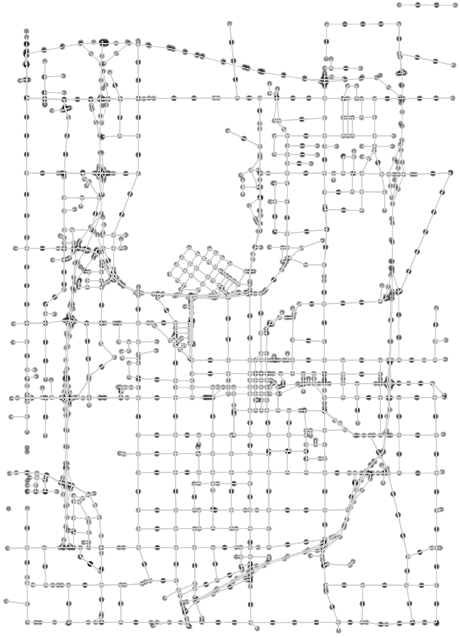}%
  \vspace{-3mm}
  \caption{The Sioux Fall Network: (a) original, (b) classic, (c) enhanced, (d) extended.}
  \label{fig:sioux-falls-network}
  \vspace{-8mm}
\end{figure}

\subsection{The Sioux Falls Network}

\subsubsection{A Brief History of the Sioux Falls Network}
%
%
The Sioux Falls network was first introduced to the research community by \citet{Morlok1973Development} in 1973 as a test case for a traffic equilibrium network.
Its widespread adoption originated with LeBlanc et al. \cite{LeBlanc1975Algorithm, LeBlanc1975efficient}, who used it as a tractable test problem to validate an efficient algorithm for solving the traffic assignment problem.
With 24 nodes, 76 links, and 552 origin-destination (O-D) pairs, the network struck a balance between simplicity and analytical relevance, enabling meaningful testing while remaining computationally tractable on the hardware of the time.
Although it was never intended to represent the real transportation system of Sioux Falls, its stylized structure and complete dataset made it a de facto benchmark for evaluating algorithms in transportation science \cite{Salim1998comparison}. 
Its accessibility and reproducibility led to widespread use across studies in traffic assignment, modal split, and network design \cite{Bar-Gera2016Sioux}.
Despite its acknowledged limitations, including arbitrary parameter choices and limited realism, the Sioux Falls network became firmly established as a standard reference in transportation research.

%
%
As transportation modeling paradigms have shifted from aggregate, static analysis to disaggregate, dynamic simulation, the network has been adapted and enriched, ensuring its continued relevance.
Recognizing the limitations of the classical Sioux Falls scenario, researchers have undertaken efforts to modernize the network for use in dynamic and agent-based modeling frameworks.
For instance, \citet{Chakirov2014Enriched} transformed the original benchmark into an enriched, agent-based scenario tailored for the MATSim platform \cite{Horni2016MultiAgent}.
This extension replaced the traditional O-D matrix with a synthetic population of over 84,000 agents, each assigned individualized daily activity schedules derived from real-world census, land-use, and building data.
The updated scenario incorporates time-varying, activity-based demand, enhanced spatial granularity, and a multi-modal transportation system, including a bus network.
Furthermore, in 2017, the MATSim Sioux Falls scenario was updated to reflect a more detailed and spatially refined network representation, expanding the topology to 1,810 nodes and 2,494 links.

\subsubsection{The Sioux Falls Network in Research Practice}
%
%
The original use of the Sioux Falls network was in static traffic assignment, where it served as a benchmark for evaluating equilibrium-based models \cite{Salim1998comparison}.
Throughout the late 1970s and 1980s, it remained a standard test case for advancing assignment algorithms.
As transportation modeling evolved, researchers extended its use to dynamic traffic assignment and time-dependent simulations, further demonstrating its adaptability \cite{Thong2021Computing}.
In parallel, the network was adopted in microsimulation and agent-based modeling, where it functioned as a virtual city for testing traveler behavior and routing under dynamic conditions \cite{Chakirov2014Enriched}.
Another major application domain is transportation network design and optimization, where Sioux Falls has long served as a canonical example for evaluating methods such as capacity expansion, link addition, and road pricing \cite{Baskan2014Harmony, Nakazato2024365day, Jiang2015Timedependent}.
More recently, as the research focus has expanded from efficiency to system robustness, the Sioux Falls network has become a key platform for studying vulnerability and resilience \cite{Hackl2019Estimation}.
In this context, it is commonly subjected to simulated disruptions to identify critical components and evaluate mitigation strategies \cite{Martinez-Pastor2021Transport, Tu2024Identifying, Xu2024Exploring}.

%
%
With the increasing adoption of data-driven methods and machine learning in transportation research, the Sioux Falls network has found new relevance in studies focused on learning and approximating traffic patterns.
One line of inquiry employs graph neural networks to emulate the traffic assignment process, mapping origin-destination (O-D) demand to link flows, without relying on iterative equilibrium solvers \cite{Liu2024Endtoend}.
Due to its small size and computational efficiency, the Sioux Falls network enables the rapid generation of synthetic datasets for training machine learning models.
It has also been used in the development of surrogate models and reinforcement learning techniques for network control and optimization \cite{Shang2025impact}.
For example, deep neural networks have been trained to approximate travel time functions or to serve as lightweight simulators, thereby accelerating optimization workflows \cite{Rahman2023DataDriven}.
In recent years, the network has also been widely utilized to evaluate the impacts of connected and autonomous vehicles \cite{Mirzahossein2024Modeling, Vitale2024Cooperative} and to test intelligent transportation system strategies \cite{Kang2022Impact, Gallo2024Combined} under controlled, reproducible conditions.

%
%
Although the Sioux Falls network was never intended to serve as a realistic representation of an actual transportation system, it continues to be widely used as a benchmark in transportation research.
Its popularity persists despite a general recognition within the community that the network is highly stylized and oversimplified.
While its simplicity offers clear advantages for algorithm testing and reproducibility, it remains unclear to what extent results derived from this benchmark generalize to more complex, real-world systems.
This raises a critical question: \textit{how representative is the Sioux Falls network, and under what conditions can conclusions drawn from it be considered reliable?}
Addressing this gap requires a rigorous mathematical framework to evaluate the structural and functional fidelity of benchmark networks relative to empirical mobility patterns.

\subsection{Stochastic Network Analysis}
%
%
Stochastic network analysis offers a principled framework for understanding complex networks by modeling their structure and dynamics as outcomes of probabilistic processes \cite{Newman2018Networks}.
Instead of analyzing a single deterministic network, this approach considers ensembles of graphs generated under specific probabilistic rules.
Thereby, random graph models, such as the Erdős-Rényi, configuration, small-world or scale-free networks \cite{Watts1998Collective, Barabasi1999Emergence, Molloy1995critical}, serve as null models that preserve selected structural properties, while randomizing others.
These models allow researchers to statistically test whether observed patterns (\emph{e.g.}, degree distributions, assortativity, clustering) could arise by chance or indicate underlying generative mechanisms such as growth, optimization, or strategic link formation.
Beyond topology, stochastic network processes capture dynamic behaviors, such as random walks, propagation, or system dynamics \cite{Masuda2017Random}, by modeling interactions as probabilistic transitions in Markovian systems.
From this perspective, stochastic network analysis offers a rigorous toolkit for hypothesis testing, model selection, and quantifying structural variability in complex systems.

\subsubsection{Higher-Order Network Representations}
%
%
While stochastic network analysis provides powerful tools for assessing the statistical significance of network structures, it typically relies on first-order representations that assume memoryless interactions, meaning that the next step in a mobility trajectory is based only on the current position.
However, the dynamics observed in real-world transportation systems often exhibit strong path dependencies and non-Markovian behavior due to inherent sequential dependencies in real-world mobility data, which arise from the intentionality of human movement from origins (\emph{e.g.}, home) toward destinations (\emph{e.g.}, work or school) \cite{Zhang2025Higherorder}.
To address these limitations, a paradigm shift in network science is emerging, adopting higher-order network representations that explicitly encode higher-order interactions such as memory effects \cite{Lambiotte2019networks}.

%
%
To represent such dependencies, higher-order networks extend conventional graph models by encoding both topological and sequential information.
A graph $\mathcal{G} = (\mathcal{V}, \mathcal{E})$ consists of a set of nodes $\mathcal{V} = {v_{1}, \dots, v_{N}}$ and a set of edges $\mathcal{E} \subseteq {e_{ij}:=(v_i, v_j) \mid \bm{A}{ij} \neq 0}$, where $\bm{A}$ denotes the adjacency matrix encoding pairwise connections \cite{Newman2018Networks}.
Flows from origin to destination nodes are represented by set of paths \( \mathcal{P} \) in the graph.
A single path \(p:=(v_0,v_1,\dots,v_\ell) \in \mathcal{P}\) is defined as an ordered sequence of nodes such that each consecutive pair of nodes corresponds to an edge in the graph.
%
%
Higher-order networks build upon this foundation by redefining nodes as ordered sequences of graph nodes, thereby preserving memory of previously visited locations.
Formally, a higher-order node is defined as
\(v^{(k)}:= \langle v_{0},v_{1},\ldots,v_{k} \rangle  = \left\{\left(v_0, v_1, \ldots, v_k\right) \mid v_i \in \mathcal{V}\right\}\),
where \(\left(v_0, v_1, \ldots, v_k\right)\) is a feasible path in \(\mathcal{G}\), and \(k\) is the ``order'' or memory length.
Transitions between these paths are represented as higher-order edges \(e_{vw}^{(k)}= (v^{(k)},w^{(k)}) = \left(\langle v_{0},v_{1}, \ldots , v_{k} \rangle , \langle v_{1}, \ldots, v_{k}, v_{k+1} \rangle\right)\) with weights \(w\) as transitions probabilities \(\bm{P}_{v,w}\) \cite{Zhang2025Connectivity}:
\begin{equation}
  \label{eq:hon}
  \text{w}(v^{(k)},w^{(k)}) := \mathbb{P} \left( v_i \mid v_{i-1}, \ldots, v_{i-k} \right) = \bm{P}_{v,w}^{(k)} = \frac{\left|\{(v_{i-k}, \ldots, v_{i-1}, v_i) \in \mathcal{P}\}\right|}{\sum_{u \in \mathcal{V}}\left|\{(v_{i-k}, \ldots, v_{i-1},  u) \in \mathcal{P}\}\right|},
\end{equation}
\noindent
where the numerator counts occurrences of the sequence $(v_{i-k},\ldots,v_i)$, and the denominator sums over all continuations $u \in \mathcal{V}$ with the same prefix $(v_{i-k},\ldots,v_{i-1})$, for normalization.

Two primary approaches exist for constructing higher-order networks that capture sequential memory: (i) topology-based methods, which derive higher-order structures from the graph's connectivity \cite{Bruijn1946combinatorial}, and (ii) path-based methods, which infer memory-dependent transitions directly from observed trajectory data \cite{Scholtes2017When}.
Despite their enriched structure, higher-order networks retain all formal properties of graphs, enabling the direct application of standard graph algorithms and analytics.
These considerations motivate a more nuanced analytical approach capable of capturing memory-dependent structures and quantifying deviations across empirical and simulated mobility networks.

\newpage

\section{Methods}{\label{sec:method}}
%
%
To evaluate the representativeness of benchmark scenarios in transportation science, one would ideally compare their outputs against high-fidelity GPS trajectory data, which captures the empirical mobility behavior of individual travelers at fine spatial and temporal resolution.
However, such data are often unavailable for benchmark networks due to privacy concerns, limited geographic coverage, or the synthetic nature of the scenarios themselves.
This necessitates the development of a statistical modeling framework that enables robust comparison across a wide range of systems, including both empirical and simulated data, as well as networks of varying sizes and complexities.
Such a framework must precisely represent mobility patterns while supporting rigorous statistical analysis.
Aligned with the principles of stochastic network analysis, our goal is to test whether statistically significant deviations arise between benchmark-based simulations and real-world mobility behavior.
While traditional stochastic models primarily capture topological characteristics, they fall short in accounting for the sequential and memory-dependent nature of movement \cite{Battiston2021physics}.
To overcome this limitation, we employ higher-order network models, recently developed at the intersection of network science and statistical physics, as a more expressive representation of complex spatiotemporal systems \cite{Lambiotte2019networks}.
These models preserve the path-dependent dynamics embedded in trajectory data while providing a principled statistical framework for cross-scenario comparison.

%
%
Although the proposed methodological framework is universally applicable, this study specifically focuses on the Sioux Falls network due to its extensive use as a standardized benchmark in transportation research (see Section~\ref{sec:background}).
To assess its representativeness, we utilize trajectory data from artificial agents navigating the Sioux Falls scenario generated using the MATSim simulation environment \cite{Horni2016MultiAgent} and compare these trajectories against openly accessible empirical GPS trajectory data.
For the simulation experiments, we consider two distinct variants of the Sioux Falls network: (i) the original (``classical'') configuration consisting of 24 nodes and 76 edges, adapted for agent-based simulations by \citet{Chakirov2014Enriched}, and (ii) an ``extended'' version featuring an enriched topology comprising 1,810 nodes and 2,494 edges.
Given the lack of open-access GPS datasets specifically collected within the Sioux Falls region, empirical trajectory data from two Chinese cities, Jinan and Shenzhen, provided by \citet{Yu2023Cityscale}, are used as representative samples of real-world urban mobility.
This setup enables us to test the ability of our framework to identify universal structural and behavioral patterns in trajectory data, as well as to quantify where and how synthetic simulations diverge from empirical observations.
Employing data from distinct geographic contexts serves as an additional test of our framework's ability to identify universal patterns in human mobility data, enabling us to determine whether observed trajectory behaviors exhibit similarities across diverse scenarios or, conversely, pinpoint the structural divergences that exist between them.

\subsection{Data Collection and Preprocessing}
To construct and analyze higher-order network representations, we utilize two types of trajectory data: synthetic agent-based simulations and real-world GPS traces.
For the simulated data, we use MATSim \cite{Horni2016MultiAgent} to generate activity-based mobility trajectories on both the classical and extended variants of the Sioux Falls network.
In both cases, the same synthetic demand, comprising 54,546 agents with individualized daily activity schedules, is applied to ensure comparability.
MATSim simulates these agents over the course of a single day, accounting for congestion dynamics, time-dependent travel times, and adaptive behavior driven by a co-evolutionary utility maximization algorithm \cite{Nagel2009Agentbased, Hackl2019Epidemic}.
The simulation yields 90,785 individual trajectories in the simplified network and 105,101 in the detailed network, each encoded as a time-ordered sequence of node IDs.

\begin{figure}[htbp]
  \vspace{-2mm}
    \centering
    \includegraphics[width=.33\textwidth]{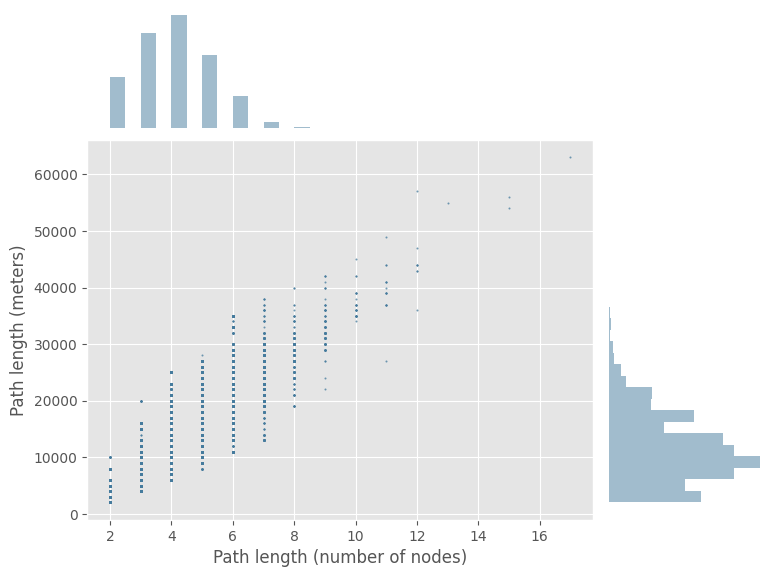}\hfill
    \includegraphics[width=.33\textwidth]{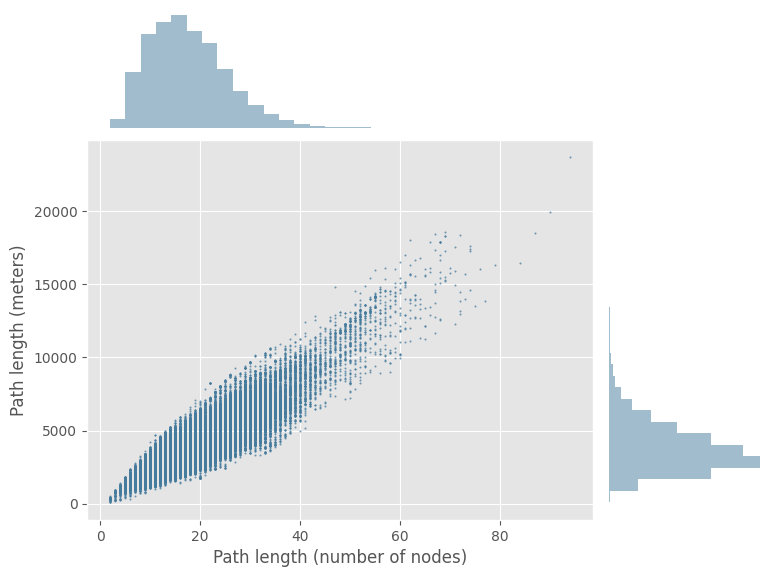}\hfill
    \includegraphics[width=.33\textwidth]{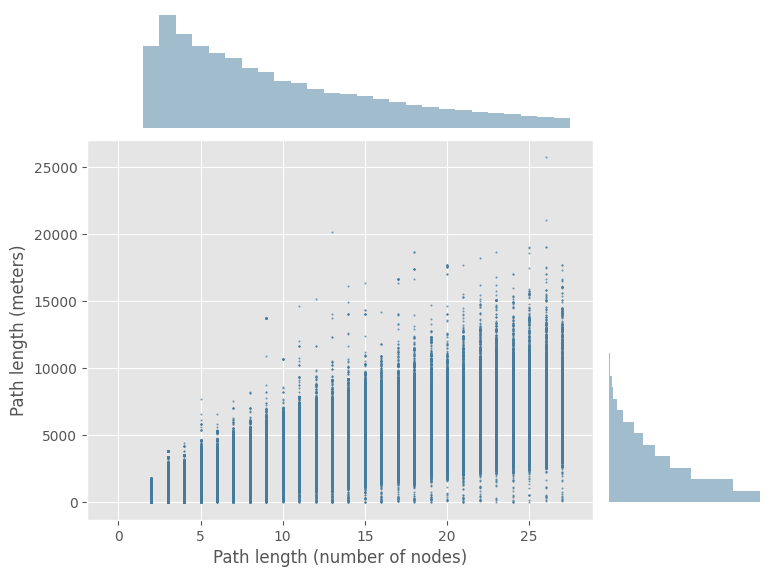}
    \vspace{-3mm}
    \caption{Path Statistics: \emph{left}: classical Sioux Falls, \emph{middle}: extended Sioux Falls, \emph{right}: Jinan}
    \label{fig:paths}
    \vspace{-4mm}
\end{figure}

In addition to the synthetic data, we incorporate real-world vehicle trajectories collected via city-scale traffic camera networks in Shenzhen and Jinan and provided by \citet{Yu2023Cityscale}.
The dataset comprises over 5 million anonymized trajectories, reconstructed using a spatiotemporal vehicle re-identification and trajectory recovery framework applied to footage from more than 3,000 traffic cameras.
All trajectory data, simulated and empirical, are extracted and preprocessed into path formats suitable for higher-order network model construction.
Descriptive statistics of the resulting path lengths across all network configurations are summarized in Figure~\ref{fig:paths}.

\subsection{Higher-Order Network Representations}
%
%
Based on the path-level data, a higher-order network model can be directly inferred from observed trajectories.
To capture the sequential dependencies inherent in mobility patterns, a fixed-length sliding window of size $k+1$ is applied to each trajectory, extracting contiguous node sequences of the form $(v_1, v_2, \ldots, v_k, v_{k+1})$.
Each subpath of length $k$, such as $(v_1, v_2, \ldots, v_k)$, is treated as a distinct higher-order node, encoding the memory of the preceding $k-1$ steps.
Directed edges between higher-order nodes are then defined by overlapping subpaths.
Specifically, an edge is added from the node $(v_1, v_2, \ldots, v_k)$ to the node $(v_2, v_3, \ldots, v_{k+1})$ if the transition $(v_k \rightarrow v_{k+1})$ is observed in the data.
This construction results in a higher-order network that faithfully encodes sequential dependencies in a computationally efficient manner, as it considers only transitions that occur in the observed trajectory data rather than the full space of possible paths.

%
A fixed $k$th-order model assumes that every transition in the network depends on the same number of previous steps.
While this simplifies modeling, it can be too rigid for real-world mobility patterns.
In practice, the influence of past movements varies, \emph{i.e.} some route choices depend only on the most recent step, while others are shaped by a longer travel history.
To capture this variation, we estimate the optimal memory length $k^{*}$ by comparing how well models of different orders explain the observed trajectories \cite{Scholtes2017When}.
Given a set of $N$ observed paths $\mathcal{P} = \{p_j\}_{j=1}^N$, we calculate the likelihood of the data under a $k$th-order model as
\begin{equation}
  \label{eq:likelihood}
  \mathcal{L}\left(\bar{M}_k \mid \mathcal{P}\right) = \prod_{j=1}^N \bar{P}^{(k)}\left(p_j\right),
\end{equation}
where $\bar{P}^{(k)}(p_j)$ is the probability of path $p_j$ under the $k$th-order model $\bar{M}_k$.
A higher-order model usually gives a better fit, but it also has more parameters, which increases the risk of overfitting \cite{Mohri2018Foundations}.
To avoid this, we apply a Likelihood Ratio Test to decide whether a higher-order model significantly improves the fit.
We compare a model of order $k$ (null model) with a model of order $k+1$ (alternative model) and compute the test statistic
\begin{equation}
\Lambda = -2 \ln \frac{\mathcal{L}\left(\bar{M}_k \mid \mathcal{P}\right)}{\mathcal{L}\left(\bar{M}_{k+1} \mid \mathcal{P}\right)}.  
\end{equation}

According to Wilks' theorem \cite{Wilks1938LargeSample}, this statistic follows a $\chi^2$ distribution with degrees of freedom $\Delta d = d(k+1) - d(k)$, which account for the number of effective parameters in each model.
We calculate $d(k)$ based on the number of valid paths in the network:
\begin{equation}
  d(k) = (|\mathcal{V}| - 1) + \sum_{i=1}^k \left[ \#\text{ paths of length } i - \#\text{ non-zero rows in } \bm{A}^i \right],  
\end{equation}
where $\bm{A}^i$ is the $i$-th power of the adjacency matrix, and non-zero rows represent valid transitions at that scale.
We then compute the $p$-value of the test using the Gamma function:
\begin{equation}
  p = 1 - \frac{\gamma\left(\frac{\Delta d}{2}, \frac{\Lambda}{2}\right)}{\Gamma\left(\frac{\Delta d}{2}\right)},
\end{equation}
where $\gamma(s, x)$ is the lower incomplete Gamma function and $\Gamma(s)$ is the complete Gamma function.
By repeating this process for increasing values of $k$, we identify the largest order for which the improvement is still statistically significant (\emph{i.e.}, $p < \epsilon$, with $\epsilon = 0.05$).
This gives us the optimal memory length $k^{*}$, which balances model accuracy with complexity and provides a robust description of the mobility dynamics.

\subsection{Higher-Order Network Analysis}
Encoding trajectory data into a higher-order network representation preserves an underlying graph structure composed of nodes and edges.
However, nodes in higher-order networks represent higher-dimensional entities capturing sequences of visited locations, enabling the explicit modeling of memory effects inherent in mobility behavior.
Despite their increased dimensionality, these higher-order networks retain the mathematical properties of graphs, allowing for the extension of conventional stochastic network analysis techniques.
To analyze the structural and functional properties of mobility behavior across benchmark and empirical networks, we employ
(i) standard network analytics to characterize features of the systems under comparison,
(ii) centrality metrics such as PageRank \cite{Page1998PageRank} to assess the influence of higher-order states within routing dynamics, and
(iii) link prediction techniques to assess how well observed trajectories can be anticipated based on inferred network structure.

\subsubsection{Network Analytics}
As a first step in evaluating the representativeness of benchmark networks, we apply standard graph-theoretic measures to quantify and compare the structural properties of higher-order networks derived from both simulated and empirical trajectory data.
Specifically, we compute global metrics such as diameter (Diam.) , network density, average shortest path length (Avg SP), and the size of the largest connected component (LCC) to characterize macroscopic connectivity patterns.
Specifically, we compute global metrics such as number of nodes and edges, diameter, average shortest path length, network density, and the size of the largest connected component to characterize macroscopic connectivity patterns.
Together, these metrics, which are explained further as the results for each are presented, provide a structural baseline for identifying deviations introduced by memory effects and serve as a foundation for cross-scenario comparison.

\subsubsection{Higher-Order PageRank}
To quantify the relative importance of nodes within mobility networks while accounting for sequential dependencies, we employ higher-order PageRank \cite{Zhang2025Connectivity}, an extension of the classical PageRank centrality measure \cite{Page1998PageRank}.
While standard PageRank assumes memoryless transitions, higher-order PageRank computes the stationary distribution of a random walk on a higher-order graph, where nodes encode subpaths of length $k$.
The resulting higher-order PageRank vector captures the long-term visitation frequency of these memory-augmented nodes.
To recover interpretable rankings over physical locations, scores are aggregated by mapping each higher-order node to its final component, yielding a first-order ranking that reflects both topological and sequential properties of trajectory data.

\subsubsection{Link Prediction}
Higher-order network models support predictive tasks such as link prediction by leveraging the sequential dependencies embedded in trajectory data.
In this framework, the probability of a path is computed using a multi-order Markov model (Equation \eqref{eq:likelihood}) that combines lower-order transitions for initial steps with higher-order transitions once sufficient context is available.
Given a current sequence of visited nodes, the link prediction is performed by selecting the node with the highest conditional transition probability based on observed path frequencies.
While prior knowledge can be incorporated to refine predictions, our implementation relies on empirical transition probabilities alone, offering a data-driven approach to anticipate future movements in mobility trajectories.
In this study, link prediction is used to assess how well different network representations capture the sequential structure of mobility trajectories.

\section{Results}{\label{sec:results}}
%
%
The results of the previously introduced analytical tasks are presented to evaluate the representativeness of benchmark networks in transportation research.
We demonstrate the applicability of higher-order network models as a rigorous framework for capturing path-dependent mobility behavior and for validating simulation outputs against empirical observations.
As a first step, we estimate the optimal memory length $k^{*}$, which quantifies the degree of sequential dependency present in mobility patterns across different datasets.
A summary of the computed structural characteristics is provided in Table~\ref{tab:summary}.
Our findings reveal that real and simulated transportation systems typically exhibit optimal orders in the range of 3 to 5, indicating that a memory-based abstraction of this dimensionality provides the best trade-off between model complexity and data fidelity.
Accordingly, for the remainder of this section, we focus our analysis to higher-order network representations up to order 5, allowing us to evaluate structural, functional, and predictive characteristics while ensuring statistical robustness.

\subsection{Structural Properties}
To assess the representativeness of benchmark networks relative to real-world mobility systems, we begin by analyzing the structural properties of higher-order network representations across memory orders $k = 1$ (standard first-order) to $k = 5$.
This analysis reveals how incorporating sequential dependencies reshapes network topology and enables a systematic comparison between benchmark-based simulations and empirical trajectory data.
We hypothesize that if benchmark networks such as the Sioux Falls model are representative of real-world transportation systems, their higher-order structural characteristics should exhibit similar patterns to those derived from empirical mobility trajectories.

\begin{table}[ht]
\centering
\footnotesize
\begin{tabularx}{0.9\linewidth}{@{}c r r r r r r r r@{}}
\toprule
\textbf{Order} & \textbf{Nodes} & \textbf{Edges} & \textbf{In-Deg.} & \textbf{Out-Deg.} & \textbf{Diam.} & \textbf{Avg SP} & \textbf{Density} & \textbf{GCC Ratio} \\
\midrule
\multicolumn{9}{l}{\textbf{Jinan} (Optimal Order 3)} \\
1 & 5843 & 13119 & 2.25 & 2.25 & 113 & 38.03 & 0.00038 & 1.00 \\
2 & 13119 & 25042 & 1.91 & 1.91 & 124 & 43.38 & 0.00015 & 0.999 \\
3 & 25042 & 41803 & 1.67 & 1.67 & 129 & 45.98 & 0.00007 & 0.995 \\
4 & 41803 & 62835 & 1.50 & 1.50 & 135 & 49.01 & 0.00004 & 0.980 \\
5 & 62835 & 86778 & 1.38 & 1.38 & 149 & 52.01 & 0.00002 & 0.965 \\
\addlinespace
\multicolumn{9}{l}{\textbf{Shenzhen} (Optimal Order 4)} \\
1 & 1141 & 2214 & 1.94 & 1.94 & 40 & 16.24 & 0.00170 & 1.00 \\
2 & 2214 & 4463 & 2.02 & 2.02 & 46 & 18.12 & 0.00091 & 1.00 \\
3 & 4463 & 8959 & 2.01 & 2.01 & 48 & 19.05 & 0.00045 & 0.999 \\
4 & 8959 & 16788 & 1.87 & 1.87 & 50 & 19.97 & 0.00021 & 0.999 \\
5 & 16788 & 28515 & 1.70 & 1.70 & 56 & 21.06 & 0.00010 & 0.997 \\
\addlinespace
\multicolumn{9}{l}{\textbf{Extended Sioux Falls} (Optimal Order 3)} \\
1 & 1646 & 2979 & 1.81 & 1.81 & 60 & 23.87 & 0.00110 & 1.00 \\
2 & 2979 & 5454 & 1.83 & 1.83 & 66 & 25.57 & 0.00061 & 1.00 \\
3 & 5454 & 7979 & 1.46 & 1.46 & 73 & 25.01 & 0.00027 & 1.00 \\
4 & 7979 & 10774 & 1.35 & 1.35 & 83 & 25.06 & 0.00017 & 1.000 \\
5 & 10774 & 13721 & 1.27 & 1.27 & 88 & 25.60 & 0.00012 & 0.998 \\
\addlinespace
\multicolumn{9}{l}{\textbf{Classical Sioux Falls} (Optimal Order 4)} \\
1 & 24 & 76 & 3.17 & 3.17 & 6 & 3.01 & 0.13768 & 1.00 \\
2 & 76 & 252 & 3.32 & 3.32 & 7 & 3.37 & 0.04421 & 1.00 \\
3 & 252 & 641 & 2.54 & 2.54 & 9 & 4.35 & 0.01013 & 1.00 \\
4 & 641 & 1081 & 1.69 & 1.69 & 15 & 6.31 & 0.00264 & 0.931 \\
5 & 1081 & 1085 & 1.00 & 1.00 & 25 & 10.34 & 0.00093 & 0.725 \\
\bottomrule
\end{tabularx}
\caption{Structural metrics of higher-order networks ($k=1$-$5$) for Jinan, Shenzhen, Extended Sioux Falls, and Classical Sioux Falls networks.}
\label{tab:summary}
\end{table}

\subsubsection{Number of Nodes and Edges.}
Higher-order networks represent memory by encoding not just locations, but sequences of locations, as nodes, effectively keeping track of where a traveler has been before. 
Thereby, the number of nodes and edges in higher-order networks increases significantly with memory order $k$.
Consequently, this results in a greater number of unique subpaths emerging in the data, as each additional step in the sequence introduces new combinations of previously visited nodes.
This expansion is both expected and desirable, as it reflects the model's capacity to represent the sequential complexity of mobility behavior and allows us to assess how accurately different networks capture the diversity of real-world travel dynamics.
In Jinan, the number of nodes increases from 5,843 at $k=1$ to 62,835 at $k=5$ (1075.3\%), and the number of edges from 13,119 to 84,778 (661.5\%).
Shenzhen shows a 1,471.3\% increase in nodes (1,141 to 16,788) and a 1,287.9\% increase in edges (2,214 to 28,515).
The extended Sioux Falls network exhibits more moderate growth, where nodes increase by 655.5\% (1,646 to 10,774) and edges by 460.59\% (2,979 to 13,721).
In contrast, the classical Sioux Falls network grows from 24 to 1,081 nodes (4,504.2\%) and from 76 to 1,085 edges (1,427.6\%).
This decline likely reflects insufficient path length to support fifth-order transitions, as the average trajectory spans only 3.82 steps.

\subsubsection{Diameter and Average Shortest Path Length}
As memory order increases, the diameter and average shortest path length of higher-order networks also grow, reflecting the increasing navigational complexity of the system.
The \textit{diameter} refers to the longest shortest path between any two nodes in the network, indicating the overall reachability, while the \textit{average shortest path length} measures the mean number of steps required to travel between all pairs of nodes.
In higher-order networks, this growth arises because transitions are restricted to empirically observed sequences, effectively removing infeasible shortcuts and making long-range connectivity more sensitive to actual path dependencies.
In Jinan, the diameter increases from 113 to 149, and the average shortest path length rises from 38.03 to 52.01.
Shenzhen follows a similar trend, with diameter growing from 40 to 56 and average path length from 16.24 to 21.06.
The extended Sioux Falls network exhibits an increases in diameter from 60 to 88, and average shortest path length from 23.87 to 25.60.
In contrast, the classical Sioux Falls network shows disproportionate growth, with diameter increasing from 6 to 25 (416.7\%) and average shortest path length from 3.01 to 10.34 (343.5\%).
This disproportionate increase suggests a lack of realistic path alternatives in the simplified network, leading to artificial disconnectedness and reduced routing flexibility as $k$ increases.

\begin{figure}[htbp]
  \vspace{-2mm}
    \centering
    \includegraphics[width=5.4cm]{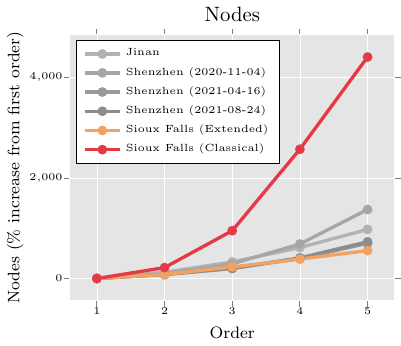}\hfill
    \includegraphics[width=5.4cm]{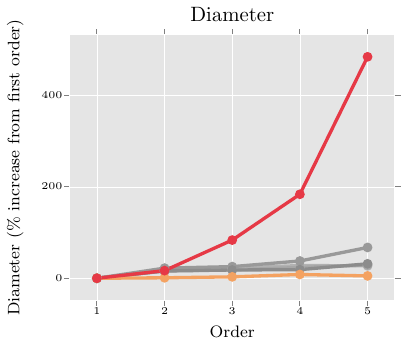}\hfill
    \includegraphics[width=5.4cm]{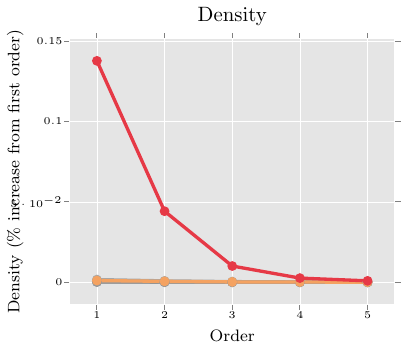}\break
    \includegraphics[width=5.4cm]{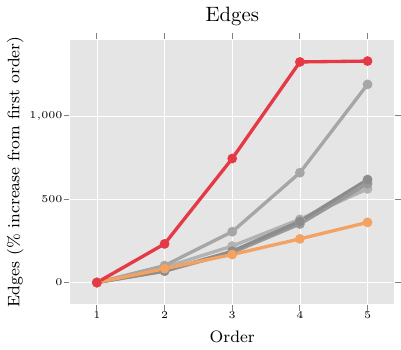}\hfill
    \includegraphics[width=5.4cm]{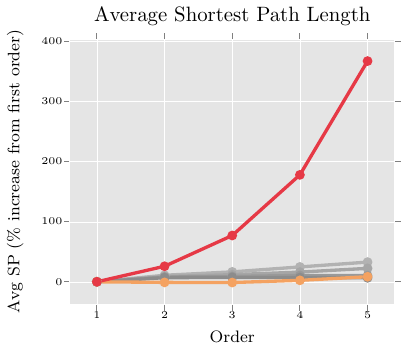}\hfill
    \includegraphics[width=5.4cm]{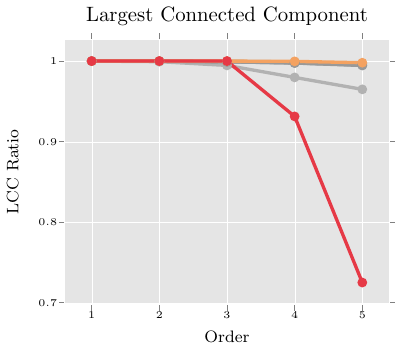}\break
    \vspace{-7mm}
    \caption{Structural Properties.}
    \label{fig:properties}
    \vspace{-4mm}
\end{figure}

\subsubsection{Density and Largest Connected Component}
Network \textit{density}, defined as the ratio of observed edges to the maximum possible number of edges, declines consistently across all datasets as the memory order increases.
This reduction occurs because higher-order networks expand the number of nodes (via memory states) much faster than the number of transitions between them, leading to sparser connectivity.
In Jinan, density drops by 94.7\%, from 0.00038 at $k=1$ to 0.00002 at $k=5$. 
Shenzhen experiences a 41.2\% decline, from 0.0017 to 0.0001.
The extended Sioux Falls network shows an 90.9\% reduction, from 0.0011 to 0.0001.
The classical Sioux Falls network exhibits the steepest decline (99.3\%) from 0.1377 to 0.00093.
Notably, its first-order density is exceptionally high compared to all other datasets, due to the small number of nodes and the relatively dense link structure in the original model.
This artificially elevated density suggests unrealistic global connectivity, which diminishes rapidly once sequential constraints are introduced and higher-order transitions are considered.
The \textit{largest connected component} (LCC), \emph{i.e.}, the largest subset of nodes in which every node is reachable from any other, remains stable across all datasets up to third order.
However, for the classical Sioux Falls network, the LCC begins to fragment at higher orders and drops to 93.1\% of the total node set at $k=4$ and further to 72.5\% at $k=5$.
This fragmentation suggests that many fifth-order transitions are not observed in the data, resulting in disconnected subgraphs and highlighting limitations in path diversity for this benchmark scenario.

\subsubsection{Similarity Analysis}
\begin{figure}[ht]
  \vspace{-2mm}
    \centering
    \includegraphics[width=\textwidth]{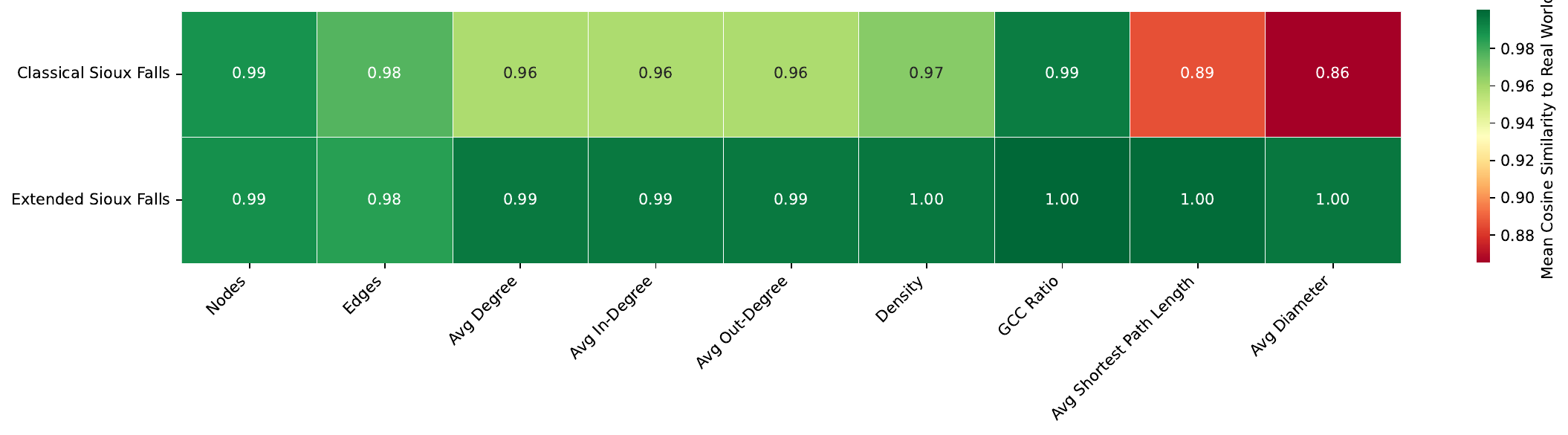}%
    \vspace{-7mm}
    \caption{Similarity Analysis.}
    \label{fig:similarity}
    \vspace{-4mm}
\end{figure}

To quantify the structural resemblance between benchmark simulations and real-world mobility patterns, we perform a similarity analysis based on the multi-order network representations introduced in this work.
Since each network property can be evaluated across memory orders $k = 1$ to $5$, we represent each scenario as a vector in a five-dimensional feature space.
Using an approach from machine learning, we compute the cosine similarity between these feature vectors to assess how closely the structural characteristics of the Sioux Falls simulations align with those derived from empirical trajectory data.
This approach allows us to capture both trends and magnitudes across memory orders in a single similarity score.
As stated in our hypothesis, a benchmark network that faithfully reflects real-world mobility behavior should exhibit high similarity scores, indicating strong structural correspondence across all orders.
The results, shown in Figure~\ref{fig:similarity}, indicate that the extended Sioux Falls network exhibits near-perfect structural alignment with empirical datasets across all metrics, with cosine similarity values above 0.99 for every feature.
In contrast, the classical Sioux Falls network, while showing high similarity in basic topological features such as node and edge count (0.99 and 0.98), diverges significantly in higher-order structural properties.
Notably, it shows substantially lower similarity in average diameter (0.86) and average shortest path length (0.89), suggesting its topology compresses or distorts network reachability and scale compared to empirical patterns.

\subsubsection{Degree Distribution}
\begin{figure}[htbp]
  \vspace{-2mm}
  \centering
    \includegraphics[width=\textwidth]{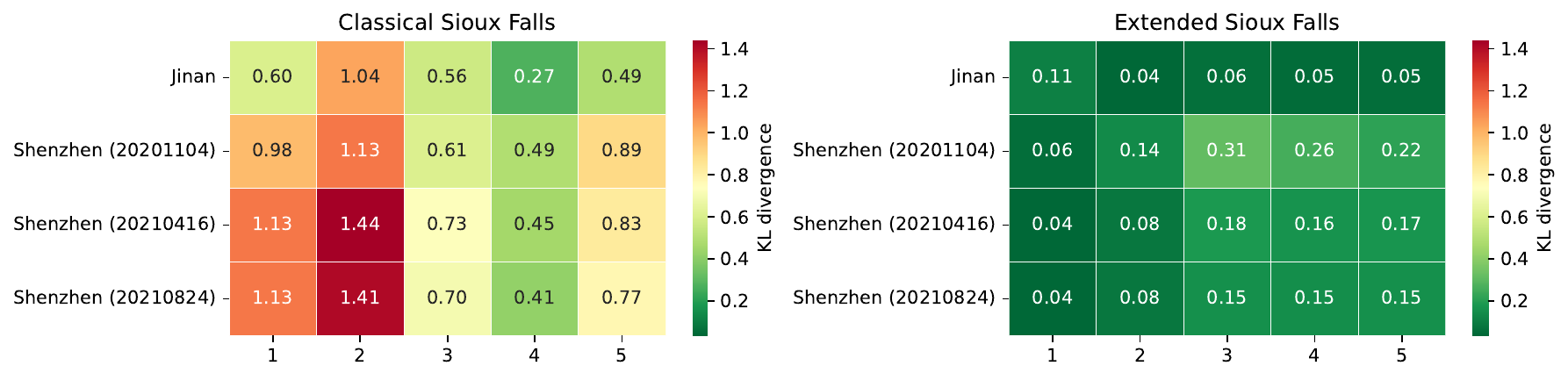}\break
    \vspace{-7mm}
    \caption{KL Divergent. Comparison between the benchmark networks degree distribution and real-world observations. Left the classical and right the extended Sioux Falls Network.}
    \label{fig:degrees}
    \vspace{-4mm}
  \end{figure}

The \textit{degree distribution} characterizes the frequency of node degrees across a network and is a fundamental descriptor of network structure as well as plays a central role in stochastic network analysis.
It provides insights into the heterogeneity of node connectivity, influences diffusion and routing dynamics, and serves as a key statistic for validating generative models.
To assess how well simulated benchmark networks replicate real-world connectivity patterns across memory orders, we compute the Kullback-Leibler (KL) divergence between degree distributions derived from Sioux Falls simulations and those observed in empirical trajectory-based networks.
As shown in Figure~\ref{fig:degrees}, the classical Sioux Falls network exhibits consistently high KL divergence values, particularly at lower memory orders (e.g., values above 1.0 for $k=2$), indicating significant dissimilarity from empirical distributions and poor fit across all datasets.
In contrast, the extended Sioux Falls network achieves uniformly low KL divergence values across all memory orders and empirical baselines, reflecting a closer match to real-world structural variability and supporting its higher representational fidelity.

\subsection{PageRank and Empirical Relevance}
To evaluate the extent to which PageRank captures empirically observed usage patterns at road intersections, we compute Kendall's Tau correlation between PageRank scores and actual intersection frequencies across increasing memory orders (the higher the better).
PageRank, a widely used centrality measure based on the stationary distribution of a random walk, reflects the likelihood of visiting each node over time. In higher-order networks, it accounts for path-dependent transitions, offering a more nuanced estimate of node importance in mobility systems.
Across all real-world and detailed simulation networks, first-order PageRank already exhibits strong alignment with empirical data, with Kendall's Tau values ranging from 0.4816 to 0.6449.
In the extended Sioux Falls network, the correlation starts at 0.5194 and improves modestly to 0.6125 at $k=3$, before declining to 0.4223 at $k=5$.
Shenzhen follows a similar trend, increasing from 0.6195 at $k=1$ to a peak of 0.7071 at $k=3$, then gradually decreasing.
Jinan also peaks at $k=3$ with a Kendall's Tau of 0.6858, followed by a more pronounced drop to 0.4650 at $k=5$.

In contrast, the classical Sioux Falls network exhibits irregular behavior.
It begins with the highest first-order correlation among all datasets ($\tau = 0.6449$), suggesting an apparently strong alignment under the memoryless model.
However, the correlation drops sharply at $k=2$ to 0.2681, a 58.4\% decline, indicating that the simplified structure does not capture second-order transitions well.
After this disruption, alignment gradually recovers, reaching 0.5145 at $k=3$ and peaking again at 0.6304 at $k=4$, before a slight decline to 0.6087 at $k=5$.
The sharp drop between $k = 1$ and $k = 2$ can be attributed to the tendency of first-order models to overestimate alignment, driven by the stylized topology and limited transition diversity of the simplified network.
This indicates that the classical Sioux Falls network exhibits a largely random structural pattern, where incorporating path memory degrades alignment, unlike in real-world networks, suggesting that sequential mobility behavior is not meaningfully captured in its topology.

\begin{figure}[ht]
  \centering
  \includegraphics{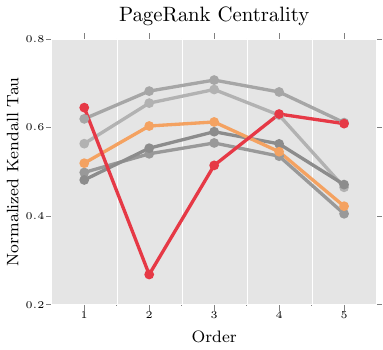}%
  \includegraphics{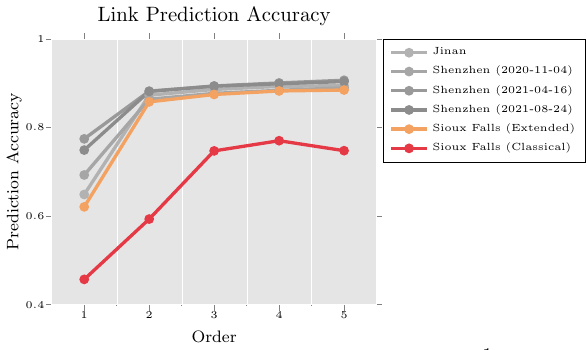}%
  \caption{PageRank analysis and link prediction accuracy for all networks.}
  \label{fig:analysis}
\end{figure}

\subsection{Link Prediction Accuracy}
To assess how well higher-order models capture the sequential structure of mobility behavior, we evaluate link prediction accuracy across memory orders $k = 1$ to $5$.
In this task, the model attempts to infer the next location in a trajectory based on the current path history, using the transition probabilities defined by the higher-order network.
Across all empirical and extended simulation networks, prediction accuracy improves sharply from first to second order, then plateaus or improves marginally, indicating that second- and third-order memory captures most of the relevant sequential dependencies.
For example, Jinan, Shenzhen, and the extended Sioux Falls network all reach accuracy levels around 0.89-0.91 at orders $k = 3$ to $5$, with optimal performance typically achieved at $k = 4$.
In contrast, the classical Sioux Falls network shows significantly lower accuracy overall and a different trajectory: while performance increases steadily up to $k = 4$ (peaking at 0.77), it drops slightly at $k = 5$.
This deviation suggests that the classical Sioux Falls network does not reflect the sequential structure of real-world mobility, as its predictive performance does not follow the expected improvement pattern seen in empirical observations.

\section{Discussion}{\label{sec:discussion}}

\subsection{Reevaluating the Sioux Falls Network as a Benchmark}
The results presented in this study reveal fundamental limitations of the classical Sioux Falls network as a benchmark for mobility modeling.
While it remains a popular testbed due to its simplicity and reproducibility, our analysis shows that its structural and functional properties deviate significantly from those observed in empirical trajectory data.
Specifically, the observed irregularities in higher-order structural and functional properties, including decreases in PageRank correlation and lower predictive accuracy, underscore the network's deficiencies in capturing sequential dependencies that characterize realistic mobility patterns.
The marked fragmentation and declining density at higher memory orders further reveal that the classical Sioux Falls topology lacks sufficient diversity in route alternatives, thus limiting its fidelity as a platform for modeling complex urban dynamics.
In contrast, the extended Sioux Falls network, which features a more detailed spatial structure, demonstrates strong alignment with empirical patterns across structural and predictive metrics.
Its performance closely mirrors that of real-world datasets such as Jinan and Shenzhen, particularly in terms of optimal memory order, predictive accuracy, and similarity of higher-order graph properties. 
These findings quantitatively confirm the classical Sioux Falls network's limited representativeness and show that a spatially enriched extension yields significantly improved alignment with real-world mobility patterns.

\subsection{Implications for Model Validation and Benchmark Design}
The application of higher-order network models offers a mathematically rigorous and empirically grounded framework for evaluating the structural and behavioral fidelity of benchmark transportation scenarios.
Traditionally, validation efforts have relied on comparing aggregate outputs such as travel times, link flows, or convergence metrics under idealized conditions, often neglecting the sequential structure and behavioral realism of agent trajectories.
By incorporating higher-order network models as diagnostic tools, we introduce a complementary framework that allows for the evaluation of benchmarks based on their ability to reproduce empirically observed path-dependent dynamics.
This approach provides a statistically grounded means to assess how well a simulation or testbed preserves memory effects, routing diversity, and structural complexity inherent to real-world systems.
Beyond evaluating benchmark scenarios, the proposed framework enables the systematic comparison of state-of-the-art transportation algorithms against empirical mobility data, facilitating a data-driven, statistical assessment of algorithmic performance and behavioral plausibility.

\subsection{Generalizability of Higher-Order Modeling Framework}
A key strength of the proposed framework lies in its broad applicability across different geographic regions, modeling paradigms, and data sources.
By encoding sequential dependencies directly from trajectory data, higher-order network models offer a unifying representation that accommodates both empirical and synthetic mobility traces, irrespective of the underlying network topology or demand model.
Our analysis demonstrates that the same modeling framework can be applied to simplified benchmarks, spatially enriched simulations, and large-scale real-world datasets, enabling direct structural and functional comparisons across these diverse scenarios.
This generalizability positions higher-order models as a powerful tool for cross-scenario benchmarking, transfer learning, and comparative infrastructure studies.
Moreover, the ability to extract interpretable patterns, such as optimal memory length, link predictability, and centrality alignment, supports robust inference even in the absence of exact spatial correspondence between datasets.
As urban mobility systems become increasingly dynamic and heterogeneous, the capacity to generalize across spatial scales and modeling resolutions is essential for building transferable, data-driven insights in transportation science.

\subsection{Limitations and Future Work}
While the proposed framework provides a rigorous foundation for evaluating the representativeness of benchmark networks, several limitations should be acknowledged.
First, the empirical comparison is constrained by the availability and geographic scope of trajectory data; due to the absence of real-world path data for Sioux Falls, empirical validation had to rely on proxy datasets from Jinan and Shenzhen.
Although these cities offer rich and diverse mobility patterns, structural and behavioral differences may exist that limit the generalizability of direct comparisons.
Second, higher-order models are subject to increased computational complexity as memory order and network size grow, which may pose challenges for real-time applications or large-scale systems unless further optimizations are implemented.
Additionally, integrating richer behavioral features, such as mode choice, temporal variation, or socio-demographic attributes, into the higher-order framework may further improve its explanatory power.

\section{Conclusion}{\label{sec:conclusion}}
This study introduced a higher-order network modeling framework to assess the representativeness of benchmark transportation networks, with a particular focus on the widely used Sioux Falls scenario.
By encoding empirical and simulated trajectory data into higher-order graphs, we systematically evaluated structural and functional properties across increasing memory orders, enabling rigorous, path-based comparisons between synthetic benchmarks and real-world mobility systems.
A summary of the computed structural characteristics is provided in Table~\ref{tab:summary}.
Our findings reveal that the classical Sioux Falls network exhibits structural artifacts and limited path diversity, resulting in declining predictive accuracy and weak alignment with empirical usage patterns when sequential dependencies are incorporated.
In contrast, a spatially extended version of the network demonstrates markedly improved performance across key metrics, including optimal memory length, link prediction accuracy, and PageRank correlation, offering a more faithful approximation of real-world mobility behavior.

The proposed framework provides a statistically grounded methodology for validating transportation models and benchmarking scenarios beyond traditional aggregate metrics.
Its applicability to both synthetic simulations and large-scale empirical datasets demonstrates the potential for establishing cross-scenario comparability and guiding the development of more realistic testbeds.
As transportation systems continue to evolve in complexity and data availability increases, integrating higher-order network representations will be essential for capturing the nuanced dynamics of traveler behavior.
Future research will extend this framework to incorporate multimodal mobility, temporal variability, and behavioral heterogeneity, further bridging the gap between simulated environments and the empirical realities of urban transportation systems.

\section{Acknowledgements}
The authors would like to thank Aleksandar Trifunovic (\url{https://github.com/akstrfn}) for creating the \verb1trbunofficial1 class document, which has been a very helpful improvement.
This manuscript benefited from the use of OpenAI's GPT-4 model for spelling and grammar checking. All scientific content, analyses, and references were developed and verified independently by the authors. The authors acknowledge the limitations of language models, including potential biases and factual inaccuracies, and confirm that no content was accepted without thorough review.

\section{Author Contributions}
The authors confirm contribution to the paper as follows: study conception and design: CZ, JH; data collection: CZ; analysis and interpretation of results: CZ, TL, JH; draft manuscript preparation: CZ, AB, TL, JH. All authors reviewed the results and approved the final version of the manuscript.

\section{Declaration of Conflicting Interests}
The authors declared no potential conflicts of interest with respect to the research, authorship, and/or publication of this article.

\section{Funding}
The authors disclosed receipt of the following financial support for the research, authorship, and/or publication of this article: This research was supported by a grant from the Energy Research Fund administered by the Andlinger Center for Energy and the Environment as well as the School of Engineering and Applied Science (SEAS) Seed Grant at Princeton University.

\bibliographystyle{trb}
\bibliography{reference}

\end{document}